\documentclass[pra,aps,onecolumn,10pt,showpacs,groupedaddress,superscriptaddress,floatfix]{revtex4-1}
\usepackage{amsmath,amsfonts,amssymb,graphics,graphicx,epsfig,color,times}
\usepackage[utf8]{inputenc}
\usepackage{color}
\usepackage{bbm, dsfont} 
\usepackage{subfigure}
\usepackage{hyperref}
\usepackage{mathrsfs}
\usepackage{verbatim}

\begin{document}

\title{Spectrally entangled biphoton state of cascade emissions from a Doppler-broadened atomic ensemble}


\author{T. H. Chang}
\affiliation{Department of Physics, National Taiwan University, Taipei 10617, Taiwan}
\author{G.-D. Lin}
\affiliation{Department of Physics, National Taiwan University, Taipei 10617, Taiwan}
\author{H. H. Jen}
\email{sappyjen@gmail.com}
\affiliation{Institute of Physics, Academia Sinica, Taipei 11529, Taiwan}




\renewcommand{\k}{\mathbf{k}}
\renewcommand{\r}{\mathbf{r}}
\newcommand{\f}{\mathbf{f}}
\def\bea{\begin{eqnarray}}
\def\eea{\end{eqnarray}}

\begin{abstract} 
We theoretically investigate the spectral property of biphoton state from the cascade emissions from a Doppler-broadened atomic ensemble. This biphoton state is spontaneously created in the four-wave-mixing process. The upper transition of the emissions lies in telecom bandwidth, which prevails in fiber-based quantum communication for low-loss transmission. We obtain the spectral property in terms of superradiant decay rates of the lower transition, excitation pulse durations, and temperature of the medium. We quantify their frequency entanglement by Schmidt decomposition and find that more entangled source can be generated with longer excitation pulses, enhanced decay rates, and significant Doppler broadening. A minimally entangled biphoton source can also be located at some optimal temperature of the atoms. This allows spectral shaping of continuous frequency entanglement, which is useful in multimode long-distance quantum communication.
\end{abstract}

\maketitle
\section{Introduction}

Quantum network \cite{Kimble2008} envisions interconnections between quantum nodes which share and transfer quantum coherence and entanglement. This distribution of entanglement or quantum communication can be achieved via efficient light-matter couplings \cite{Hammerer2010} and coherent transfer of light \cite{Gisin2007}. For fiber-based light transfer, quantum communication to long distance can be attainable if the telecom bandwidth of light is exploited, since it has the lowest transmission loss. Except for rare-earth atoms \cite{McClelland2006, Lu2010} or erbium-doped crystal \cite{Lauritzen2010}, which have telecom ground state transitions, conventional alkali-metal atomic ensembles do not directly access the telecom bandwidth from their ground states. Nevertheless, the telecom photons can be obtained via atomic cascade transitions (upper transition) \cite{Chaneliere2006} or telecom-wavelength conversion \cite{Radnaev2010, Jen2010}. The highly correlated telecom and infrared photons are spontaneously generated from four-wave mixing with two weak laser driving fields. The infrared transition (lower transition), on the other hand, is perfectly suitable for quantum storage as a genuine quantum node. Recent progress of telecom photon generation includes frequency conversions in trapped ions \cite{Walker2018, Bock2018}, a nitrogen-vacancy center in diamond \cite{Dreau2018}, or nonlinear microring resonator \cite{Li2016}, and enhanced emissions from single erbium ions in a silicon nanophotonic cavity \cite{Dibos2018}. 

In addition, high communication capacity can be feasible and encoded in either discrete \cite{Wang2018_Pan} or continuous degrees of freedom \cite{Braunstein2005}, allowing high-dimensional control and manipulation of quantum information. Continuous entanglement, to name a few, involve spatial \cite{Grad2012}, time-energy \cite{Branning1999, Law2000, Parker2000}, transverse momenta \cite{Law2004, Moreau2014}, and orbital angular momenta of light \cite{Arnaut2000, Mair2001, Molina2007, Dada2011, Fickler2012, Fickler2016}. Using atomic cascade transitions, continuous frequency entanglement \cite{Jen2012-2} can be spectrally shaped \cite{Jen2016a, Jen2016b} by frequency or phase modulations and manipulated by projecting multiphoton entangled states which are created from cascaded atomic ensembles \cite{Jen2017}. 

With the optimal telecom bandwidth in fiber-based quantum communication and potentially high capacity of frequency entanglement of the cascade emissions, we further consider a spectrally-entangled biphoton source from a thermal atomic system. This can be realized in conventional vapor cells \cite{Phillips2001,Balabas2010} which have several advantages over cold atoms in their scalability and convenient implementation with no need of trapping and cooling. For thin vapor cells \cite{Sarkisyan2004,Keaveney2012}, they can even provide strong confinement of the thermal atoms and thus the flexibility to control the cell thickness, while suffer from low optical depth. There are wide fundamental studies operated in thermal atoms, for example, including electromagnetically induced transparency and slow light properties \cite{Phillips2001,Kocharovskaya2001, Bae2010, Klein2011, Shu2016, Wang2018}, radio-frequency electrometry \cite{Sedlacek2012, Kumar2017} using Rydberg transitions \cite{Kubler2010}, atom interferometry \cite{Biedermann2017}, and quantum memory \cite{Hosseini2011, Hosseini2011_2} in scalable photonic quantum network \cite{Borregaard2016}.

In this paper, we study the biphoton states via cascade transitions from a Doppler-broadened atomic ensemble, and investigate their spectral entanglement properties. The biphoton states are spontaneously created from the cascade emissions driven by two weak laser fields under four-mixing condition. We derive the spectral function of the biphoton state with Doppler broadenings. For two excitation platforms of co- and counter-propagating schemes, we are able to numerically obtain its Von Neumann entropy by Schmidt decomposition, with dependences on driving field durations, superradiant decay constant of the infrared photon, and temperature of the thermal gas. Our study puts forward a potential application in high capacity quantum communication using spectral shaping and manipulation. 
  
\begin{figure}[t]
\centering
\includegraphics[width=8.0cm,height=6cm]{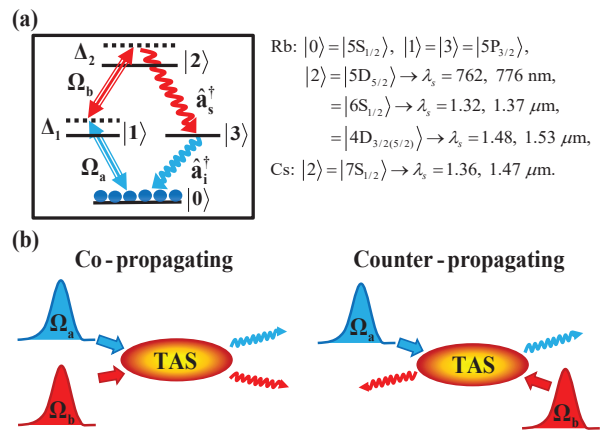}
\caption{Atomic cascade transitions and two schemes of generating biphoton states. (a) In the four-level atomic configuration of alkali-metal atoms, two weak laser fields of Rabi frequencies $\Omega_a$ and $\Omega_b$ excite the atoms into the upper state $|2\rangle$ with detunings $\Delta_1$ and $\Delta_2$. The cascade emissions $\hat a^\dag_s$ (signal) and $\hat a^\dag_i$ (idler) are generated subsequently in the spontaneous emission process. Some upper levels of rubidium and cesium atoms are shown for telecommunication wavelengths between $1.3$-$1.5$ $\mu$m. (b) Schematic co- and counter-propagating schemes for biphoton generation from a thermal atomic system (TAS).}\label{fig1}
\end{figure}

\section{Theoretical model}

\subsection{Spectral function of Doppler-broadened biphoton state}

Before including Doppler effect from a thermal atomic system (TAS), we first review how we derive the biphoton state from a cold atomic ensemble driven by two weak laser fields under four-wave mixing (FWM) condition. As shown in Fig. \ref{fig1}(a), two classical fields (with Rabi frequencies $\Omega_a$ and $\Omega_b$) drive the system from the ground state $|0\rangle$ via excited states $|1\rangle$, $|2 \rangle$, $|3\rangle$, into two spontaneously emitted photons $\hat a_s$, $\hat a_i$. The Hamiltonian in interaction picture can be written as \cite{Jen2016a, Jen2017}
\begin{eqnarray}
V_{\rm I}&=&-\sum_{m=1,2}\Delta_m\sum_{\mu=1}^N|m\rangle_\mu\langle m|-\sum_{m=a,b}\left(\frac{\Omega_m}{2}\hat{P}_m^\dag+{\rm h.c.}\right)\nonumber\\
&&-i\sum_{m=s,i}\bigg\{\sum_{\k_m,\lambda_m}g_m\hat{a}_{\k_m,\lambda_m}\hat{Q}_m^\dag e^{-i\Delta\omega_m t}-{\rm h.c.}\bigg\},\label{H}
\end{eqnarray} 
where $\hbar$ $=$ $1$, and $\lambda_m$ indicates the polarization of photons. We denote the signal and idler photon coupling constants as $g_{s(i)}$ which have encapsulated $(\epsilon_{\k_m,\lambda_m}\cdot\hat{d}_m^*)$ with the polarization direction $\epsilon_{\k_m,\lambda_m}$ of quantized bosonic fields $\hat{a}_{\k_m,\lambda_m}$ and the unit direction $\hat{d}_m$ of dipole operators. The detunings are defined as $\Delta_1$ $=$ $\omega_a$ $-$ $\omega_1$ and $\Delta_2$ $=$ $\omega_a$ $+$ $\omega_b$ $-$ $\omega_2$ with atomic level energies $\omega_{1,2,3}$. The upper excited state $|2\rangle$ allows a telecom wavelength within $1.3$-$1.5$ $\mu$m if 6S$_{1/2}$ and 4D$_{3/2(5/2)}$, or 7S$_{1/2}$ levels are considered \cite{Chaneliere2006} for rubidium or cesium atoms respectively. Various dipole operators are defined as $\hat{P}_a^\dag$ $\equiv$ $\sum_\mu|1\rangle_\mu\langle 0|e^{i\k_a\cdot\r_\mu}$, $\hat{P}_b^\dag$ $\equiv$ $\sum_\mu|2\rangle_\mu\langle 1|e^{i\k_b\cdot\r_\mu}$, $\hat{Q}_s^\dag$ $\equiv$ $\sum_\mu|2\rangle_\mu\langle 3|e^{i\k_s\cdot\r_\mu}$, and $\hat{Q}_i^\dag$ $\equiv$ $\sum_\mu|3\rangle_\mu\langle 0|e^{i\k_i\cdot\r_\mu}$. Central frequencies and wave vectors of these fields are $\omega_{a(b),s(i)}$ and $\k_{a(b),s(i)}$ respectively, where we further define $\Delta\omega_s$ $\equiv$ $\omega_s$ $-$ $\omega_2$ $+$ $\omega_3$ $-$ $\Delta_2$ and $\Delta\omega_i$ $\equiv$ $\omega_i$ $-$ $\omega_3$.

From the Hamiltonian of Eq. (\ref{H}), we can formulate self-consistent Schr\"{o}dinger equations assuming there is only one atomic excitation. And as such, we express the state function as 
\bea
|\psi(t)\rangle&=&\mathcal{E}(t)|0,{\rm vac}\rangle+\sum^N_{\mu=1} A_\mu(t)|1_\mu,vac\rangle
+\sum^N_{\mu=1} B_\mu(t)|2_\mu,vac\rangle+\sum^N_{\mu=1}\sum_{s}C^\mu_s(t)|3_\mu,1_{\k_s,\lambda_s}\rangle\nonumber\\
&&+\sum_{s,i}D_{s,i}(t)|0,1_{\k_s,\lambda_s},1_{\k_i,\lambda_i}\rangle,
\eea
where $s(i)$ in the summation denotes $k_{s(i)}$ and $\lambda_{s(i)}$ respectively, collective single excitation states are $|m_\mu\rangle$ $\equiv$ $|m_\mu\rangle|0\rangle^{\otimes N-1}_{\nu\neq\mu}$ with $m$ $=$ $1$, $2$, $3$, and $|{\rm vac}\rangle$ represents the vacuum photon state. The state function involves the states with a FWM cycle where a collectively excited atom goes through the intermediate, upper excited states, and then back to the ground state with signal and idler emissions. The assumption of single excitation space is valid when weak laser fields are applied and satisfy $\sqrt{N}|\Omega_a|/\Delta_1\ll 1$. This requirement also complies with the condition of adiabatic elimination of the levels $|1\rangle$ and $|2\rangle$ in the excitation process.

Using Schr\"odinger equation $i\hbar\frac{\partial}{\partial t}|\psi(t)\rangle$ $=$ $V_{\rm I}(t)|\psi(t)\rangle$, we are able to derive self-consistent coupled equations of motion. With the steady-state solutions, the ground state probability amplitude $\mathcal{E}(t)$ is almost unity, and  
\begin{eqnarray}
A_\mu(t)&\approx& -\frac{\Omega_a(t)}{2\Delta_1}e^{i\k_a\cdot\r_\mu},\\
B_\mu(t)&\approx& \frac{\Omega_a(t)\Omega_b(t)}{4\Delta_1\Delta_2}e^{i(\k_a+\k_b)\cdot\r_\mu},
\end{eqnarray}
where the atom adiabatically follows the driving fields. We further consider a symmetrical single excitation state, $(\sqrt{N})^{-1}$ $\sum_{\mu=1}^N$ $e^{i(\k_a+\k_b-\k_s)\cdot\r_\mu}|3\rangle_\mu|0\rangle^{\otimes N-1}$, which contributes to the biphoton generation most significantly in large $N$ limit, and therefore this leads to the probability amplitude of the biphoton state $|1_{\k_s},1_{\k_i}\rangle$ \cite{Jen2012-2, Jen2016a, Jen2017},
\begin{eqnarray}
D_{s,i}(t)= g_{i}^{\ast}g_{s}^{\ast}\sum_{\mu=1}^Ne^{i\Delta\k\cdot\r_{\mu}}\int_{-\infty}^{t}\int_{-\infty}^{t^{\prime}}dt^{\prime\prime}dt^{\prime}
e^{i\Delta\omega_{i}t^{\prime}}e^{i\Delta\omega_{s}t^{\prime\prime}} \frac{\Omega_{a}(t'')\Omega_{b}(t'')}{4\Delta_{1}\Delta_{2}}e^{\left(-\Gamma_{3}^N/2+i\delta\omega_{i}\right)(t^{\prime}-t^{\prime\prime})},\label{Dsi2}
\end{eqnarray}
where FWM condition $\sum_{\mu=1}^N e^{i\Delta\k\cdot\r_{\mu}}$ guarantees the phase-matched and highly correlated biphoton state when $\Delta\k$ $\equiv$ $\k_{a}$ $+$ $\k_{b}$ $-$ $\k_{s}$ $-$ $\k_{i}$ $\rightarrow$ $0$. The superradiant decay constant \cite{Dicke1954} of the idler photon \cite{Chaneliere2006, Jen2012} can be quantified as $\Gamma_{3}^{\rm N}$ $=$ $(N\bar{\mu}+1)\Gamma_{3}$ with an intrinsic decay rate $\Gamma_3$ and geometrical constant $\bar{\mu}$ \cite{Rehler1971}. This constant in general depends on the atomic density and light-matter interacting volume, which therefore can be easily modified by controlling the thickness of vapor cell or its density. The relevant collective frequency shift \cite{Friedberg1973, Scully2009} is denoted as $\delta\omega_{i}$, which originates from resonant dipole-dipole interaction from rescattering spontaneous emissions \cite{Lehmberg1970}. This shift can be absorbed into idler central frequency, which is generally order of kilohertz \cite{Jen2015}.

To proceed, we assume Gaussian pulse excitations, $\Omega_{a,b}(t)$ $=$ $\tilde{\Omega}_{a,b}e^{-t^{2}/\tau^{2}}/(\sqrt{\pi}\tau)$, with the pulse duration $\tau$ to characterize the effect of excitations on spectral properties of the biphoton state. $\tilde{\Omega}_{a,b}$ represents the pulse area, and we obtain $D_{si}$ from Eq. (\ref{Dsi2}),
\begin{eqnarray}
D_{si}(\Delta\omega_{s},\Delta\omega_{i})&=&\frac{\tilde{\Omega}_{a}\tilde{\Omega}_{b}g_{i}^{\ast}g_{s}^{\ast}}{4\Delta_{1}\Delta_{2}}\frac{\sum_{\mu}e^{i\Delta\k\cdot\r_{\mu}}}{\sqrt{2\pi}\tau}f(\omega_s,\omega_i),\\
f(\omega_s,\omega_i)&\equiv&
\frac{e^{-(\Delta\omega_{s}+\Delta\omega_{i})^{2}\tau^{2}/8}}{\frac{\Gamma_{3}^{N}}{2}-i\Delta\omega_{i}},\label{f}
\end{eqnarray}
where $f(\omega_s,\omega_i)$ has a Lorentzian distribution with a spectral width $\Gamma_{3}^{N}/2$ for the idler photon and a joint Gaussian distribution for signal and idler ones. The joint Gaussian profile maximizes the state generation near $\Delta\omega_{s}$ $+$ $\Delta\omega_{i}$ $=$ $0$. This shows energy-conserving correlated biphoton state driven by respective laser fields, that is, $\omega_s+\omega_i=\omega_a+\omega_b$.

Finally, to include the Doppler broadening in the spectral function, we average $f(\omega_s,\omega_i)$ of Eq. (\ref{f}) with a Maxwell-Boltzmann distribution for some temperature $T$ of the thermal atoms. We then obtain  
\bea
f_D(\omega_s,\omega_i)=\int_{-\infty}^\infty f(\omega_s-k_sv,\omega_i\mp k_i v)\frac{e^{-v^2/(2\sigma^2)}}{\sqrt{2\pi}\sigma}dv,\label{fd}
\eea
where $\sigma\equiv\sqrt{k_BT/m}$ with the mass of the atom $m$ and the Boltzmann constant $k_B$. The sign of $\mp k_i v$ indicates the co- or counter-propagating schemes respectively, which we will investigate in details later. Here one-dimensional average of Maxwell-Boltzmann distribution is kept intact due to FWM condition. From Eq. (\ref{fd}) and considering the copropagating scheme, we further obtain
\bea
f_D(\omega_s,\omega_i)&=&\frac{-i}{\sqrt{2\pi}\sigma k_i}e^{-\tau^2(1-b)(\Delta\omega_s+\Delta\omega_i)^2/8}e^{-A^2}\left[\pi{\rm Erfi}\left(A\right)+i\pi\right],\label{fD}\\
A&\equiv&\sqrt{\frac{\tau^2}{8b}}
\frac{\left[b(k_i/\bar{k}_{si})\Delta\omega_s+(bk_i/\bar{k}_{si}-1)\Delta\omega_i-i\Gamma_3^N/2\right]}{k_i/\bar{k}_{si}},\label{A}
\eea  
where $b\equiv\bar{k}_{si}^2/[\bar{k}_{si}^2+4/(\sigma\tau)^2]$, $\bar{k}_{si}\equiv k_s+k_i$ with $k_{s,i}=|\k_{s,i}|$, and Erfi is imaginary error function with a complex argument $A$. This complex error function can be re-expressed by $(2/\sqrt{\pi})e^{A^2}D(A)$ with the Dawson function $D(A)$ which we implement numerically in mathematical software. The special function arises due to the form of integration in Eq. (\ref{fd}), which involves a multiplication of Lorentzian and Gaussian profiles, and it is related to Faddeeva or Voigt functions \cite{Abramowitz1965}.

\subsection{Spectral analysis}

The spectral properties and continuous frequency entanglement of the Doppler-broadened biphoton state can be analyzed by Schmidt decomposition. Here we review and introduce Schmidt decomposition in continuous frequency spaces \cite{Law2000}. For some polarizations $\lambda_{s}$ and $\lambda_{i}$, we express the Doppler-broadened biphoton state $|\Psi\rangle$ with a spectral function $f_D(\omega_{s},\omega_{i})$,
\begin{equation}
|\Psi\rangle=\int f_D(\omega_{s},\omega_{i})\hat{a}_{\lambda_{s}}^{\dag}(\omega_{s})\hat{a}_{\lambda_{i}}^{\dag}(\omega_{i})|0\rangle d\omega
_{s}d\omega_{i}.
\end{equation}
The quantification of entanglement in the above biphoton state can be determined in the Schmidt bases where the state vectors can be written as
\begin{eqnarray}
|\Psi\rangle&=&\sum_{n}\sqrt{\lambda_{n}}\hat{b}_{n}^{\dag}\hat{c}_{n}^{\dag}|0\rangle,\\
\hat{b}_{n}^{\dag}&\equiv&\int\psi_{n}(\omega_{s})\hat{a}_{\lambda_{s}}^{\dag}(\omega_{s})d\omega_{s},~
\hat{c}_{n}^{\dag}\equiv\int\phi_{n}(\omega_{i})\hat{a}_{\lambda_{i}}^{\dag}(\omega_{i})d\omega_{i},
\end{eqnarray}
where $\hat{b}_{n}^{\dag}$ and $\hat{c}_{n}^{\dag}$ are effective photon creation operators with eigenmodes $\psi_{n}$ and $\phi_{n}$ respectively. $\lambda_n$'s are the eigenvalues and the probabilities for $n$th eigenmode. These eigen solutions are obtained by
\begin{eqnarray}
&&\int K_{1}(\omega,\omega^{\prime})\psi_{n}(\omega^{\prime})d\omega^{\prime}  =\lambda_{n}\psi_{n}(\omega),\\
&&\int K_{2}(\omega,\omega^{\prime})\phi_{n}(\omega^{\prime})d\omega^{\prime}  =\lambda_{n}\phi_{n}(\omega),
\end{eqnarray}
where 
\begin{eqnarray}
&&K_{1}(\omega,\omega^{\prime}) \equiv\int f_D(\omega,\omega_{1})f_D^{\ast}(\omega^{\prime},\omega_{1})d\omega_{1},\\
&&K_{2}(\omega,\omega^{\prime}) \equiv\int f_D(\omega_{2},\omega)f_D^{\ast}(\omega_{2},\omega^{\prime})d\omega_{2}. 
\end{eqnarray}
The above $K_{1,2}$ are the kernels for one-photon spectral correlations \cite{Law2000, Parker2000}. The orthogonality of eigenmodes is satisfied that $\int\psi_{i}(\omega)$$\psi_{j}^*(\omega)d\omega$ $=$ $\delta_{ij}$, $\int\phi_{i}(\omega)$$\phi_{j}^*(\omega)d\omega$ $=$ $\delta_{ij}$, and the normalization of quantum state demands $\sum_{n}\lambda_{n}$ $=$ $1$.

The Von Neumann entropy of entanglement $S$ in such Schmidt bases can then be calculated as
\begin{equation}
S=-\sum_{n=1}^{\infty}\lambda_{n}\textrm{log}_2\lambda_{n}.\label{entropy}
\end{equation}
$S$ is vanishing when $\lambda_{1}=1$, which indicates of a separable or non-entangled state. For more than
one Schmidt numbers $\lambda_n$, the entropy $S>0$, meaning nonvanishing bipartite entanglement. Finite $S$ indicates that $f_D(\omega_{s},\omega_{i})$ cannot be factorized as a multiplication of two separate spectral functions, that is $g_D(\omega_{s})h_D(\omega_{i})$.

\section{Copropagating scheme}

\begin{figure}[t]
\centering
\includegraphics[width=11.0cm,height=6cm]{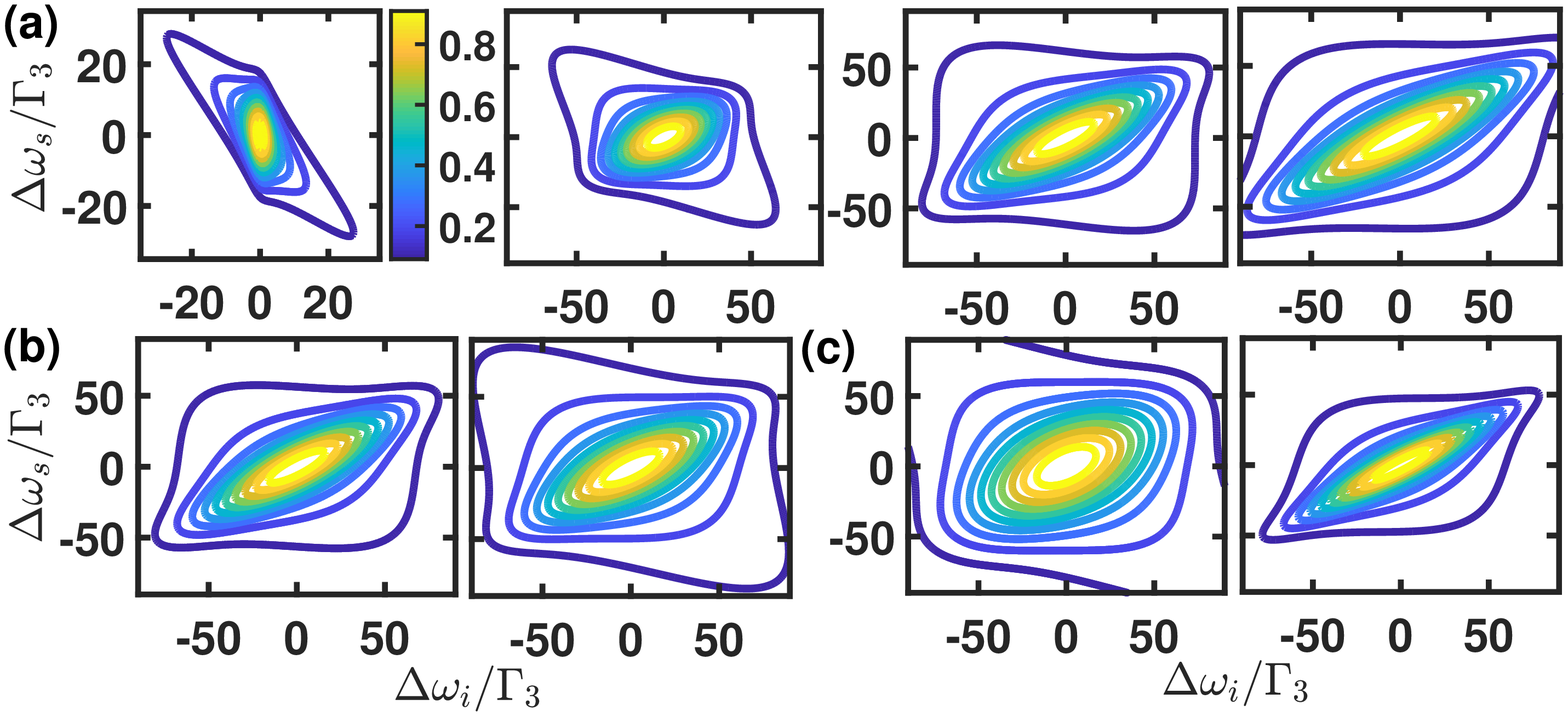}
\caption{Doppler-broadened spectral functions $|f_D(\omega_s,\omega_i)|$ in frequency spaces. (a) Spectral functions for various temperatures, $T\rightarrow 0$, $T=100$, $300$, and $500$ K, from left to right. The other parameters are $\Gamma_3^N/\Gamma_3=5$ and $\Gamma_3\tau=0.25$. Comparing the case of $T=300$ K, we show $|f_D(\omega_s,\omega_i)|$ for superradiant decay constants of $\Gamma_3^N/\Gamma_3=2.5$ and $10$ respectively in (b), and pulse durations of $\Gamma_3\tau=0.125$ and $0.5$ respectively in (c). All $|f_D(\omega_s,\omega_i)|$'s are normalized and share the same color bar in (a).}\label{fig2}
\end{figure}

In the copropagating scheme of two-photon excitation, we expect of significant Doppler broadening in the spectral function $f_D(\omega_s,\omega_i)$. This suggests that the temperature of the thermal atoms can act as a controllable knob over spectral entanglement. Below and throughout the article we take $\lambda_i=795$ nm, $\lambda_s=1.32~\mu$m, and $\Gamma_3=2\pi\times 5.8$ MHz of $D1$ transition as an example in rubidium atoms. 

In Fig. \ref{fig2}, we demonstrate the effects of temperature, excitation pulse duration, and superradiant decay constant on the biphoton spectral function. As the system temperature increases in Fig. \ref{fig2}(a), the spectrum distributes toward the anti-diagonal direction. This trend can be seen in Eq. (\ref{A}) where copropagating scheme makes $(bk_i/\bar{k}_{si}-1)<0$ at high temperature as $b\rightarrow 1$. Therefore, the spectral distribution of $\Delta\omega_s$ has an opposite sign to $\Delta\omega_i$, and Eq. (\ref{fD}) favors the distribution of anti-correlated signal and idler photons. 

\begin{figure}[t]
\centering
\includegraphics[width=11.0cm,height=6cm]{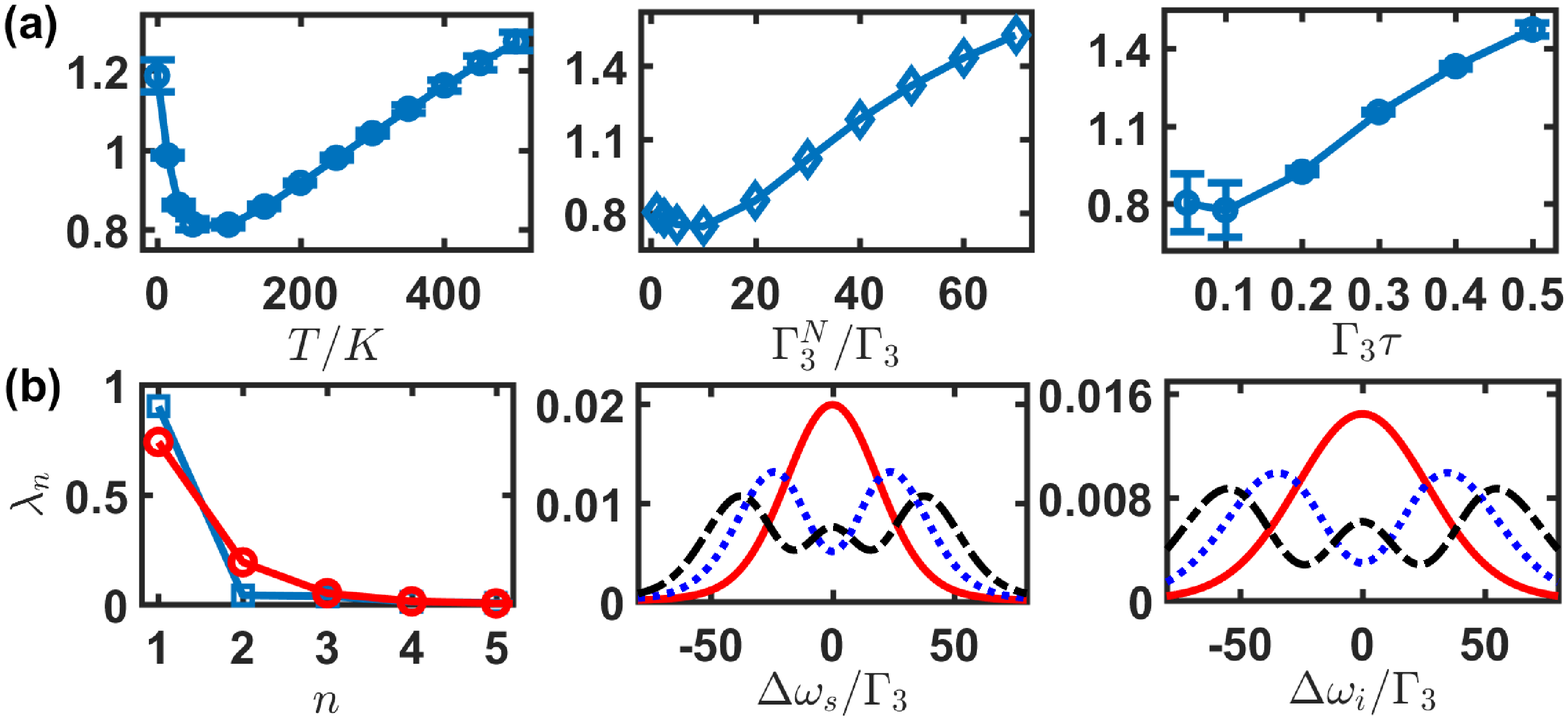}
\caption{Entropy of entanglement and Schmidt decomposition for copropagating scheme. (a) The bipartite entanglement for different temperature of thermal gas, superradiant decay constants ($T=100$ K), and excitation pulse durations. Error bars show the fitting with $95\%$ confidence level, except the middle panel calculated with a range of $\Delta\omega_{s,i}=\pm 150\Gamma_3$. (b) The most significant five Schmidt numbers for $T=50$ ($\square$) and $500$ K ($\circ$), and corresponding first three mode functions (solid, dots, and dash-dots) of $|\psi_s|^2$ and $|\phi_i|^2$ respectively at $T=500$ K. All other parameters are fixed at $T=300$ K, $\Gamma_3^N/\Gamma_3=5$, or $\Gamma_3\tau=0.25$, unless specifically stated.}\label{fig3}
\end{figure}

This transition of distributions from diagonal to anti-diagonal directions further shows the competition between two energy scales of $\Gamma_3^N$ and $k_B T$. In Fig. \ref{fig2}(b), at room temperature, the effect of $\Gamma_3^N$ tends to distribute the spectral function along the energy-conserving axis (diagonal direction), that is $\Delta\omega_s+\Delta\omega_i=0$. This is more transparent when we look at the low $T$ limit in Fig. \ref{fig2}(a) where $b\rightarrow 0$, and the Gaussian distribution $e^{-\tau^2(\Delta\omega_s+\Delta\omega_i)^2/8}$ from Eq. (\ref{fD}) dominates. As shown in Fig. \ref{fig2}(c), smaller pulse duration allows broader spectral ranges, which makes the distribution more symmetric to the center frequencies of signal and idler photons ($\Delta\omega_s=\Delta\omega_i=0$), while longer pulses tighten the spectral ranges, making the distribution more anti-diagonal. This dependence on pulse durations will also reflect on the entanglement property we discuss later in Fig. \ref{fig3}.  

We further analyze the bipartite entanglement in continuous frequency spaces by Schmidt decomposition. In a Doppler-broadened medium, the entanglement is convex to the dependences of temperature or superradiant decay constants, as shown in Fig. \ref{fig3}(a). It reaches the minimum near the transition from diagonal to anti-diagonal spectral distributions. The minimum indicates of a more symmetric spectral function with the lowest entanglement, while when $T$ or $\Gamma_3^N$ increases, we have a more entangled biphoton source. For the dependence of pulse durations, the entanglement increases almost linearly within our parameter regimes. We calculate the entanglement by using Eq. (\ref{entropy}), and find the asymptotic entropy of entanglement at infinite spectral ranges by fitting it with a function $a(1-e^{-bR})$. $R$, $a$, and $b$ respectively represent the limited ranges we apply, fitted asymptotic $S$ at infinite ranges, and slope of convergence. The numerical calculations of special functions in Eq. (\ref{fD}) are limited by $64$-digit computer double-precision, thus allowing limited spectral ranges for fitting. As expected, $S$ approaches $a$ as more data points are included for fitting, while we note that for small $T$ and $\tau$, the fitting deviations are relatively large due to the limited ranges.     
   
In Fig. \ref{fig3}(b), we compare two cases of thermal atoms with a low or high temperature. The more entangled biphoton state at high $T$ reflects on the slow decay of Schmidt eigenvalues $\lambda_n$. Large entanglement means more modes involved in frequency spaces, which can be used for encoding information in quantum communication. As a demonstration, we show first three mode functions for the case of high $T$. Broader spectral widths for the idler photon compared to the signal one can be seen in the spectral weights in Fig. \ref{fig2}(a). The number of the resonances corresponds to the order of signal and idler modes, and they resemble the Hermite-Gaussian modes in space due to the nature of Gaussian distribution in the spectral function.    

\section{Counter-propagating scheme}

For the counter-propagating scheme, the sign of $\k_s$ is opposite to $\k_i$, which leads to $|\bar k_{si}|<k_i$. Thus, we expect of less Doppler effect on the spectral function. However, it is still interesting to explore the spectral properties in such system, which provides an alternative and flexible manipulation of spectral entanglement.  

\begin{figure}[t]
\centering
\includegraphics[width=11.0cm,height=6cm]{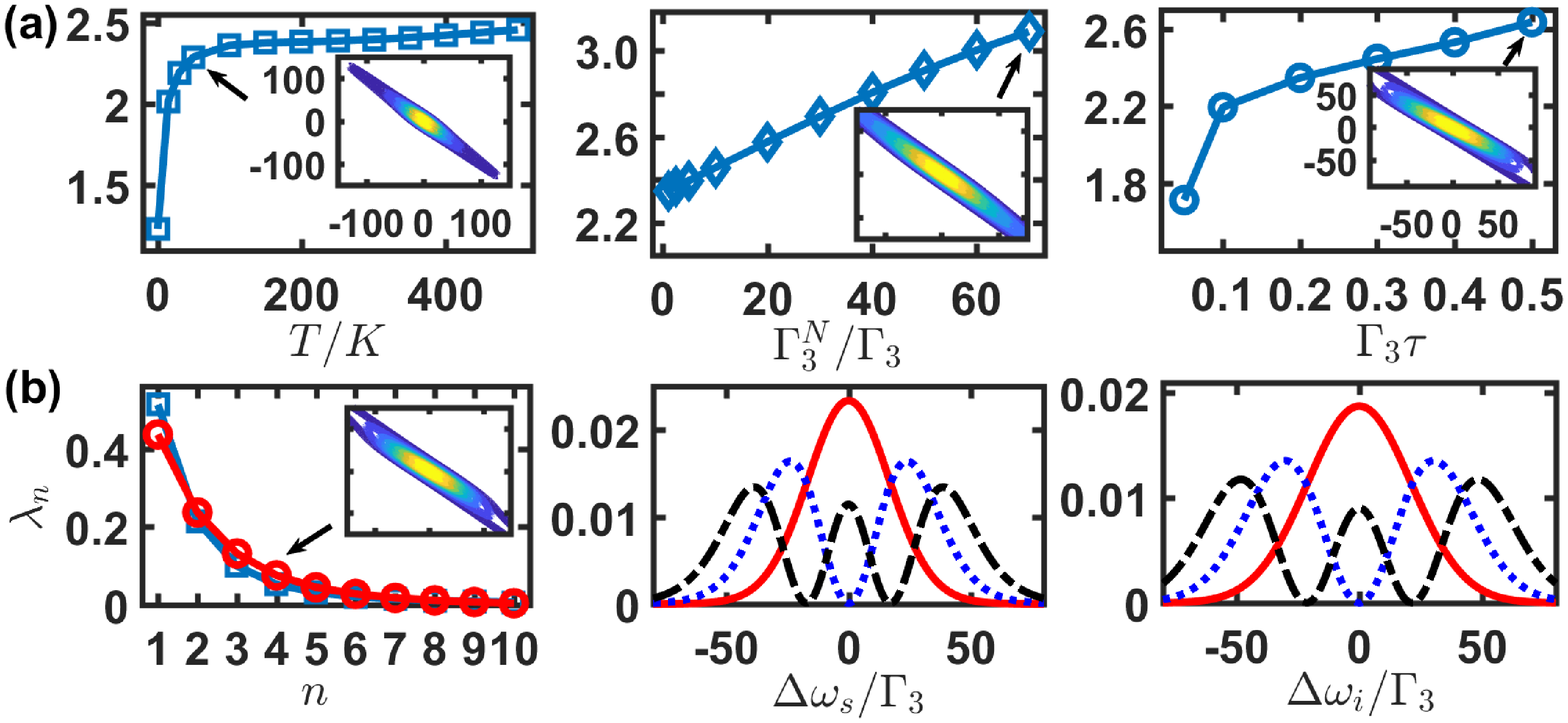}
\caption{Asymptotic entropy of entanglement $S$ and Schmidt decomposition for counter-propagating scheme. (a) The bipartite entanglement as a dependence of the temperature of thermal gas, superradiant decay constants, and excitation pulse durations. (b) The most significant ten Schmidt numbers for $T=50$ ($\square$) and $500$ K ($\circ$), and corresponding first three mode functions (solid, dots, and dash-dots) of $|\psi_s|^2$ and $|\phi_i|^2$ respectively at $T=500$ K. All other parameters are fixed at $T=300$ K, $\Gamma_3^N/\Gamma_3=5$, or $\Gamma_3\tau=0.25$. The insets in (a) and (b) are spectral functions $|f_D(\omega_s,\omega_i)|$ with the same parameter ranges as the leftmost one of (a), unless specified. Arrows indicate the specific parameters for the spectral functions.}\label{fig4}
\end{figure}

In Fig. \ref{fig4}(a), we again plot the bipartite entanglement with the dependences of three main parameters, similar to Fig. \ref{fig3}(a). Various insets in the panels show the spectral functions for specific parameter regimes indicated by the arrows. As $T$ increases, the spectral distribution tends to have more weights along the diagonal direction, and thus allows a more entangled biphoton source as shown in Fig. \ref{fig3}(b). The entropy of entanglement saturates for large $T$, which indicates a limitation on increasing $S$ by tuning the temperature of the system. The $S$ increases linearly as $\Gamma_3^N$ increases or at large $\tau$. By contrast, the entanglement of biphoton state created from the counter-propagating scheme does not have the valley-like dependence in the copropagating case. Therefore, the counter-propagating scheme seems to provide less control over low entropy of entanglement. Nevertheless, the saturation of temperature dependence allows for more stable control over entanglement, which is thus more resistant to thermal fluctuations of the system. Similarly, we plot first three mode functions in Fig. \ref{fig4}(b) at high temperature. Comparing Fig. \ref{fig3}(b), we find that the modes are more distinctive in the resonance peaks, making the spectral shaping of modes more advantageous.

\section{Discussion and Conclusion}

Continuous frequency entanglement empowers the quantum capacity for efficient and secure quantum communication. The advantage lies not only in potentially infinite modes being exploited, but also the robust and versatile coherent control through optical elements. With convenient platforms of vapor cells, Doppler-broadened and spectrally-correlated biphoton source allows for entanglement control and mode manipulations. Furthermore, using spectral shaping \cite{Lukens2014} by multiplexing multiple atomic ensembles \cite{Jen2016a, Jen2016b} enables encoding quantum information to realize quantum algorithms with Hadamard gates \cite{Lukens2014}. This promises a compact, stable, and controllable implementation of entangled photons, which are fundamental to scalable quantum information processing \cite{Lukens2017, Kues2017}.

In conclusion, we have investigated the spectral entanglement of the biphoton state from a thermal atomic system. We derive the spectral function by including the Doppler effect, and consider co- and counter-propagating excitation schemes to generate such biphoton state from atomic cascade configurations. We are able to spectrally manipulate and shape the entanglement property by controlling the temperature of the atoms, excitation pulse durations, or superradiant decay constant of the lower transition. More entangled biphoton source can be created for higher temperature, decay rates, or longer pulse durations. On the other hand, minimally entangled or nearly pure biphoton state can be located by tuning the temperature or decay rates of the system in the copropagating scheme. Our study paves the way toward multimode quantum information processing by engineering spectral modes of the photons, and potentially enables efficient high-dimensional quantum communication via telecom bandwidth.    

\section*{Funding}

Ministry of Science and Technology (MOST) of Taiwan (MOST-106-2112-M-001-005-MY3, 106-2811-M-001-130, 107-2811-M-001-1524).

\section*{Acknowledgments}

GDL thanks the support from MOST of Taiwan under Grant No. 105-2112-M-002-015-MY3 and National Taiwan University under Grant No. NTU-106R891708. We are also grateful for the support of NCTS ECP1 (Experimental Collaboration Program).



\begin{thebibliography}{99}
\bibitem{Kimble2008} H. J. Kimble, ``The quantum internet,'' Nature {\bf 493}, 1023--1030 (2008). 
\bibitem{Hammerer2010} K. Hammerer, A. S. S{\o}rensen, and E. S. Polzik, Quantum interface between light and atomic ensembles. Rev. Mod. Phys. {\bf 82}, 1041 (2010).
\bibitem{Gisin2007} N. Gisin and R. Thew, ``Quantum communication,'' Nat. Photonics {\bf 1}, 165--171 (2007).
\bibitem{McClelland2006} J. J. McClelland and J. L. Hanssen, ``Laser Cooling without Repumping: A
Magneto-Optical Trap for Erbium Atoms,'' \prl {\bf 96}, 143005 (2006).
\bibitem{Lu2010} M. Lu, S. H. Youn, and B. L. Lev, ``Trapping Ultracold Dysprosium: A Highly
Magnetic Gas for Dipolar Physics'' \prl {\bf 104}, 063001 (2010).
\bibitem{Lauritzen2010} B. Lauritzen, J. Min\'a\ifmmode \check{r}\else \v{r}\fi{}, H. de Riedmatten, M. Afzelius, N. Sangouard, C. Simon, and N. Gisin, ``Telecommunication-wavelength solid-state memory at
the single photon level,'' \prl {\bf 104}, 080502 (2010). 
\bibitem{Chaneliere2006} T. Chaneli\`{e}re, D. N. Matsukevich, S. D. Jenkins, T. A. B. Kennedy, M. S. Chapman, and A. Kuzmich, ``Quantum telecommunication based on atomic cascade transitions,'' \prl {\bf 96}, 093604 (2006).
\bibitem{Radnaev2010} A. G. Radnaev, Y. O. Dudin,	R. Zhao, H. H. Jen,	S. D. Jenkins, A. Kuzmich, and T. A. B. Kennedy, ``A quantum memory with telecom-wavelength conversion,'' Nat. Phys. {\bf 6}, 894--899 (2010).
\bibitem{Jen2010} H. H. Jen and T. A. B. Kennedy, ``Efficiency of light-frequency conversion in an atomic ensemble,'' \pra {\bf 82}, 023815 (2010).
\bibitem{Walker2018} T. Walker, K. Miyanishi, R. Ikuta, H. Takahashi, S. V. Kashanian, Y. Tsujimoto, K. Hayasaka, T. Yamamoto, N. Imoto, and M. Keller, ``Long-distance single photon transmission from a trapped ion
via quantum frequency conversion,'' \prl {\bf 120}, 203601 (2018).
\bibitem{Bock2018} M. Bock, P. Eich, S. Kucera, M. Kreis, A. Lenhard, C. Becher, and J. Eschner, ``High-fidelity entanglement between a trapped ion and a telecom photon via quantum frequency conversion,'' Nat. Commun. {\bf 9}, 1998 (2018).
\bibitem{Dreau2018} A. Dr\'eau, A. Tcheborateva, A. E. Mahdaoui, C. Bonato, and R. Hanson, ``Quantum frequency conversion of single photons from a nitrogen-vacancy center in diamond to telecommunication wavelengths," Phys. Rev. Applied {\bf 9}, 064031 (2018).
\bibitem{Li2016} Q. Li, M. Davanc, and K. Srinivasan, ``Efficient and low-noise single-photon-level frequency conversion interfaces using silicon nanophotonics,'' Nat. Photonics {\bf 10}, 406--414 (2016).
\bibitem{Dibos2018} A. M. Dibos, M. Raha, C. M. Phenicie, and J. D. Thompson, ``Atomic source of single photons in the telecom band,'' \prl {\bf 120}, 243601 (2018).
\bibitem{Wang2018_Pan} X.-L. Wang, Y.-H. Luo, H.-L. Huang, M.-C. Chen, Z.-E. Su, C. Liu, C. Chen, W. Li, Y.-Q. Fang, X. Jiang, J. Zhang, L. Li, N.-L Liu, C.-Y Lu, and J.-W. Pan, ``18-Qubit entanglement with six photons' three degrees of freedom,'' \prl {\bf 120}, 260502 (2018).
\bibitem{Braunstein2005} S. L. Braunstein and P. van Loock, ``Quantum information with continuous variables,'' Rev. Mod. Phys. {\bf 77}, 513--577 (2005).
\bibitem{Grad2012} A. Grodecka-Grad, E. Zeuthen, and A. S. S\o rensen, ``High-capacity spatial multimode quantum memories based on atomic ensembles,'' \prl {\bf 109}, 133601 (2012).
\bibitem{Branning1999} D. Branning, W. P. Grice, R. Erdmann, and I. A. Walmsley, ``Engineering the indistinguishability and entanglement of two photons,'' \prl {\bf 83}, 955--958 (1999).
\bibitem{Law2000} C. K. Law, I. A. Walmsley, and J. H. Eberly, ``Continuous frequency entanglement: Effective finite Hilbert space and entropy control,'' \prl {\bf 84}, 5304--5307 (2000).
\bibitem{Parker2000} S. Parker, S. Bose, and M. B. Plenio, ``Entanglement quantification and purification in continuous-variable systems,'' \pra {\bf 61}, 032305 (2000).
\bibitem{Law2004} C. K. Law and J. H. Eberly, ``Analysis and interpretation of high transverse entanglement in optical parametric down conversion,'' \prl {\bf 92}, 127903 (2004).
\bibitem{Moreau2014} P.-A. Moreau, F. Devaux, and E. Lantz, ``Einstein-Podolsky-Rosen paradox in twin images,'' \prl {\bf 113}, 160401 (2014).
\bibitem{Arnaut2000} H. H. Arnaut and G. A. Barbosa, ``Orbital and intrinsic angular momentum of single photons and entangled pairs of photons generated by parametric down-conversion,'' \prl {\bf 85}, 286--289 (2000).
\bibitem{Mair2001} A. Mair, A. Vaziri, G. Weihs, and A. Zeilinger, ``Entanglement of the orbital angular momentum states of photons,'' Nature {\bf 412}, 313--316 (2001).
\bibitem{Molina2007} G. Molina-Terriza, J. P. Torres, and L. Torner, ``Twisted photons,'' Nat. Phys. {\bf 3}, 305--310 (2007).
\bibitem{Dada2011} A. C. Dada, J. Leach, G. S. Buller, M. J. Padgett, and E. Andersson, ``Experimental high-dimensional two-photon entanglement and violations of generalized Bell inequalities,'' Nat. Phys. {\bf 7}, 677--680 (2011).
\bibitem{Fickler2012} R. Fickler, R. Lapkiewicz, W. N. Plick, M. Krenn, C. Schaeff, S. Ramelow, A. Zeilinger, `` Quantum entanglement of high angular momenta,'' Science {\bf 338}, 640--643 (2012).
\bibitem{Fickler2016} R. Fickler, G. Campbell, B. Buchler, P. K. Lam, and A. Zeilinger, ``Quantum entanglement of angular momentum states with quantum numbers up to $10,010$,'' Proc. Natl. Acad. Sci. U.S.A. {\bf 113}, 13642-13647 (2016).
\bibitem{Jen2012-2} H. H. Jen, ``Spectral analysis for cascade-emission-based quantum communication in atomic ensembles,'' J. Phys. B: At. Mol. Opt. Phys. {\bf 45}, 165504 (2012).
\bibitem{Jen2016a} H. H. Jen and Y.-C. C, ``Spectral shaping of cascade emissions from multiplexed cold atomic ensembles,'' \pra {\bf 93}, 013811 (2016).
\bibitem{Jen2016b} H. H. Jen, ``Entropy of entanglement in the continuous frequency space of the biphoton state from multiplexed cold atomic ensembles,'' J. Phys. B: At. Mol. Opt. Phys. {\bf 49}, 035503 (2016).
\bibitem{Jen2017} H. H. Jen, ``Cascaded cold atomic ensembles in a diamond configuration as a spectrally
entangled multiphoton source,'' \pra {\bf 95}, 043840 (2017).
\bibitem{Phillips2001} D. F. Phillips, A. Fleischhauer, A. Mair, R. L. Walsworth, and M. D. Lukin, ``Storage of light in atomic vapor,'' \prl {\bf 86}, 783--786 (2001).
\bibitem{Balabas2010} M. V. Balabas, K. Jensen, W. Wasilewski, H. Krauter, L. S. Madsen, J. H. M\"{u}ller, T. Fernholz, and E. S. Polzik, ``High quality anti-relaxation coating material for alkali atom vapor cells,'' Opt. Express {\bf 18}, 5825--5830 (2010)
\bibitem{Sarkisyan2004} D. Sarkisyan, T. Varzhapetyan, A. Sarkisyan, Yu. Malakyan, A. Papoyan, A. Lezama, D. Bloch, and M. Ducloy, ``Spectroscopy in an extremely thin vapor cell: Comparing the cell-length dependence in fluorescence and in absorption techniques,'' \pra {\bf 69}, 065802 (2004).
\bibitem{Keaveney2012} J. Keaveney, A. Sargsyan, U. Krohn, I. G. Hughes, D. Sarkisyan, and C. S. Adams, ``Cooperative Lamb shift in an atomic vapor layer of nanometer thickness,'' \prl {\bf 108}, 173601 (2012).
\bibitem{Kocharovskaya2001} O. Kocharovskaya, Y. Rostovtsev, and M. O. Scully, ``Stopping light via hot atoms,'' \prl {\bf 86}, 628--631 (2001).
\bibitem{Bae2010} I.-H. Bae, H. S. Moon, M.-K. Kim, L. Lee, and J. B. Kim, ``Transformation of electromagnetically induced transparency into enhanced absorption with a standing-wave coupling field in an Rb vapor cell,'' Opt. Express {\bf 18}, 1389--1397 (2010).
\bibitem{Klein2011} M. C. Klein and M. Hohensee, D. F. Phillips, and R. L. Walsworth, ``Electromagnetically induced transparency in paraffin-coated vapor cells,'' \pra {\bf 83}, 013826 (2011).
\bibitem{Shu2016} C. Shu, P. Chen, T. K. A. Chow, L. Zhu, Y. Xiao, M.M.T. Loy, and S. Du, ``Subnatural-linewidth biphotons from a Doppler-broadened hot atomic vapour cell,'' Nat. Commun. {\bf 7}, 12783 (2016).
\bibitem{Wang2018} G. Wang, Y.-S. Wang, E. K. Huang, W. Hung, K.-L. Chao, P.-Y. Wu, Y.-H. Chen, and Ite A. Yu, ``Ultranarrow-bandwidth filter based on a thermal EIT medium,'' Sci. Rep. {\bf 8}, 7959 (2018).
\bibitem{Sedlacek2012} J. A. Sedlacek, A. Schwettmann, H. K\"ubler, R. L\"ow, T. Pfau, and J. P. Shaffer, ``Microwave electrometry with Rydberg atoms in a vapour cell using bright atomic resonances,'' Nat. Phys. {\bf 8}, 819--824 (2012).
\bibitem{Kumar2017} S. Kumar, H. Fan, H. K\"ubler, A. J. Jahangiri, and J. P. Shaffer, ``Rydberg-atom based radio-frequency electrometry using frequency modulation spectroscopy in room temperature vapor cells,'' Opt. Express {\bf 25}, 8625--8637 (2017).
\bibitem{Kubler2010} H. K\"ubler, J. P. Shaffer, T. Baluktsian, R. L\"ow, and T. Pfau, ``Coherent excitation of Rydberg atoms in micrometre-sized atomic vapour cells,'' Nat. Photonics {\bf 4}, 112--116 (2010).
\bibitem{Biedermann2017} G. W. Biedermann, H. J. McGuinness, A. V. Rakholia, Y.-Y. Jau, D. R. Wheeler, J. D. Sterk, and G. R. Burns, ``Atom interferometry in a warm vapor,'' \prl {\bf 118}, 163601 (2017). 
\bibitem{Hosseini2011} M. Hosseini, G. Campbell, B. M. Sparkes, P. K. Lam, and B. C. Buchler, ``Unconditional room-temperature quantum memory,'' Nat. Phys. {\bf 7}, 794--798 (2011).
\bibitem{Hosseini2011_2} M. Hosseini, B.M. Sparkes, G. Campbell, P.K. Lam, and B.C. Buchler, ``High efficiency coherent optical memory with warm rubidium vapour,'' Nat. Commun. {\bf 2}, 174 (2011).
\bibitem{Borregaard2016} J. Borregaard, M. Zugenmaier, J. M. Petersen, H. Shen, G. Vasilakis, K. Jensen, E. S. Polzik, and A. S. S{\o}rensen, ``Scalable photonic network architecture based on motional averaging in room temperature gas,'' Nat. Commun. {\bf 7}, 11356 (2016).
\bibitem{Dicke1954} R. H. Dicke, Coherence in spontaneous radiation processes, Phys. Rev {\bf 93}, 99 (1954).
\bibitem{Jen2012} H. H. Jen, ``Positive-$P$ phase-space-method simulation of superradiant emission from a cascade atomic ensemble,'' \pra {\bf 85}, 013835 (2012).
\bibitem{Rehler1971} N. E. Rehler and J. H. Eberly, ``Superradiance,'' \pra {\bf 3}, 1735--1751 (1971).
\bibitem{Friedberg1973} R. Friedberg, S. R. Hartmann, J. T. Manassah, ``Frequency shifts in emission and absorption by resonant systems of two-level atoms,'' Phys. Rep. {\bf 7}, 101--179 (1973).
\bibitem{Scully2009} M. O. Scully, ``Collective Lamb shift in single photon Dicke superradiance,'' \prl {\bf 102}, 143601 (2009).
\bibitem{Lehmberg1970} R. H. Lehmberg, ``Radiation from an N-atom system. I. General formalism,'' \pra {\bf 2}, 883--888 (1970).
\bibitem{Jen2015} H. H. Jen, ``Superradiant cascade emissions in an atomic ensemble via four-wave mixing,'' Annals of Physics {\bf 360}, 556--570 (2015).
\bibitem{Abramowitz1965} M. Abramowitz and I. A. Stegun, ``Handbook of Mathematical Functions with Formulas, Graphs, and Mathematical Tables,'' (Dover Publications, Inc., New York, 1964). 
\bibitem{Lukens2014} J. M. Lukens, A. Dezfooliyan, C. Langrock, M. M. Fejer, D. E. Leaird, and A. M. Weiner, ``Orthogonal spectral coding of entangled photons,'' Phys. Rev. Lett. {\bf 112}, 133602 (2014).
\bibitem{Lukens2017} J. M. Lukens and P. Lougovski, ``Frequency-encoded photonic qubits for scalable quantum information processing,'' Optica {\bf 4}, 8--16 (2017).
\bibitem{Kues2017} M. Kues, C. Reimer, P. Roztocki, L. R. Cort\'{e}s, S. Sciara, B. Wetzel, Y. Zhang, A. Cino, S. T. Chu, B. E. Little, D. J. Moss, L. Caspani, J. Aza\~{n}a, and R. Morandotti, ``On-chip generation of high-dimensional entangled quantum states and their coherent control,'' Nature {\bf 546}, 622--626 (2017).
\end{thebibliography}
\end{document}